# Perturbative photon flux generated by high-frequency relic gravitational waves and utilization of them for their detection

Fangyu Li[1], R. M. L. Baker, Jr.[2], and Zhenya Chen[1]

[1.] *Department of Physics, Chongqing University, Chongqing 400044, P. R. China*
[2.] *GRAWAVE® LLC, 8123 Tuscany Avenue, Playa del Rey, California 90293, USA*

## Abstract

**Since 1978 relic and primordial background gravitational waves have been of increasing scientific interest. Both the quintessential inflationary models (QIM) and the pre-big bang scenario (PBBS) predict high energy density regions of relic gravitons in the microwave band (gravitational wave (GW) frequencies, $v_g$ of $10^9$-$10^{10}$Hz). There exist corresponding metric perturbations of the relic gravitational waves (GWs) in the region of approximately *h*~$10^{-30}$-$10^{-32}$. A detector for these GWs is described in which we measure the perturbative photon flux (PPF) or signal generated by such high-frequency relic GWs (HFRGWs) via a coupling system of fractal membranes and a Gaussian beam (GB) passing through a static magnetic field. It is found that under the synchro-resonance condition in which the frequency of the GB is set equal to the frequency of the expected HFRGWs (h~$2.00\times10^{-31}$, $v_g=10^{10}$Hz in the QIM and h~$6.32\times10^{-31}$, $v_g=10^{10}$Hz in the PBBS) may produce the PPFs of ~$4.04\times10^2 s^{-1}$ and ~$1.27\times10^3 s^{-1}$ in a surface of $10^{-2} m^2$ area at the waist of the GB, respectively. The relatively weak first-order PPF, directed at right angles to the expected HFRGWs, is reflected by fractal membrane and the resulting reflected PPF (signal) exhibits a very small decay in transit to the detector (tunable microwave receiver) compared with the much stronger background photon flux, which allows for detection of the reflected PPF with signal to background noise ratios greater than one at the distance of the detector. We also discuss the system's selection capability and directional sensitivity for the resonance components from the stochastic relic GW background. The resolution of tiny difference between the PPFs generated by the relic GW's in the QIM and in the PBBS may be established and will be of cosmological significance.**

------------------------------------------------------------------

1. E-mail address: fangyuli@cqu.edu.cn
2. E-mail address: robert.baker.jr@comcast.net

# 1. Introduction

Unlike usual celestial gravitational waves (GWs) having low frequencies that are often a small fraction of a Hz, the relic GWs in the microwave band ($\sim 10^9 - 10^{11}$ Hz), predicted by the quintessential inflationary models (QIM) [1-3] and the pre-big bang scenario (PBBS) [4-6], form high-frequency random signals, and because of their weakness and very high-frequency properties, they are quite different from the low-frequency GWs. Although the relic GWs has not yet been detected, we can be sure that the Earth is bathed in the sea of the relic GWs. Since 1978 such relic and primordial background GWs have been of ever increasing scientific interest as many researches have shown [7-9]. In the past few years high-frequency relic gravitational wave (HFRGW) detectors have been fabricated at Birmingham University, England [10] and INFN Genoa, Italy [11, 12]. These two types of electromagnetic (EM) detectors may be promising for the detection of the HFRGWs in the GHz band in the future, but currently their sensitivities are orders of magnitude less than what is required. The EM detection of the GWs described herein is based upon the well-respected GW theory first put forth by Gertsenshtein in 1962 [13] and is the subject of many scientific papers since that time [14-21]. On the other hand, based on high-dimensional (bulk) spacetime theories, it has been shown [22, 23] that all the familiar matter fields are constrained to live on our brane world, while gravity is free to propagate in the extra dimensions, and the high-frequency GWs (HFGWs, i.e., high-energy gravitons) would be more capable of carrying energy from our 3-brane world than lower-frequency GWs. It is noted that propagation of the HFGWs may be a unique and effective way for exchanging energy and information between two adjacent parallel brane worlds (especially the brane worlds containing thinking beings! [24-27]). Moreover, if the pre-big bang scenario is correct, then the relic GWs would be an almost unique window from which one can look back into the universe before the Big Bang [5,6,28]. Although these theories and scenarios may be controversial, whether or not they have included a fatal flaw remains to be determined.

In this paper we shall discuss some ideas for selection and detection of the HFRGWs with the predicted typical parameters $\nu_g \sim 10^{10}\,\text{Hz}$ (10GHz) and $h \sim 10^{-30} - 10^{-31}$ [1-6, 9]. This paper includes following five parts: (1) the general approximate form of the HFRGWs in the GHz band; (2) the EM resonant system to the HFRGWs, i.e., the coupling system of the fractal membranes and a Gaussian beam (GB) passing through a static magnetic field; (3) the EM resonant response to the HFRGWs and some numerical estimations; (4) splitting and pumping of the perturbative photon flux (PPF) from the background photon flux and (5) an estimate of the signal-to-noise ratio that is to be expected.

It is found that:

(1) under the synchro-resonance condition (when the frequency $\nu_e$ of the GB is tuned to the frequency $\nu_g$ of the HFRGWs), using the inverse Gertsenshtein effect and full-reflection or full-transmission properties of the new-type fractal membranes [29-31] to the photon flux in the GHz band, then the first-order perturbative photon flux, (PPF) or signal produced by the HFRGWs in the GHz band, when reflected by the fractal membranes, will exhibit much less decay rate than the background photon flux (BPF, which has a typical Gaussian decay rate) and can be detected. The PPFs contain only the first power of the HFRGW amplitude $h$ but not the second power $h^2$.

(2) the instantaneous values of the PPF may reach up to $\sim 2.02 \times 10\,\text{s}^{-1}$-$1.27 \times 10^3\,\text{s}^{-1}$ in a surface, of $10^{-2}\,\text{m}^2$ area;

(3) it is possible to obtain large signal-to-noise ratios at the distance of the detector, or tuned microwave receiver, from the reflecting fractal membrane.

(4) the size of the entire system can be limited to typical laboratory dimension (1~10m); and

(5) the system will have the capability to display directivity of the resonant components from the stochastic relic GW background.

## 2. The HFRGWs in the GHz band

It is well known that the generic form of each polarization component of the HFRGWs in the GHz band can be given by [1]

$$h = \frac{\mu(\eta)}{a}[\exp(i\mathbf{k}\cdot\mathbf{x}) + \exp(-i\mathbf{k}\cdot\mathbf{x})], \qquad (1)$$

with

$$\mu(\eta) = A_1(k)\exp(-ik\eta) + A_2(k)\exp(ik\eta), \qquad (2)$$

where $a = a(\eta)$ is the cosmology scale factor and $\eta$ is the conformal time. For the resonant response in laboratory, we should use the intervals of laboratory time (i.e., $cdt = a(\eta)d\eta$) and laboratory frequency [7]. In this case, Eqs. (1) and (2) can be described as the metric perturbation to the background spacetime, i.e.,

$$h(\mathbf{x},t) = A(k_g)/a(t)\exp[i(\mathbf{k}_g\cdot\mathbf{x} - \omega_g t)] + B(k_g)/a(t)\exp[i(\mathbf{k}_g\cdot\mathbf{x} + \omega_g t)], \qquad (3)$$

where $A/a$ and $B/a$ are the stochastic values of the amplitudes of the relic GW in the laboratory frame of reference. Equation (3) can be seen as the approximate form of each "monochromatic" polarization component of the HFRGWs in the GHz band. However, Eq. (3) shows that the stochastic relic GWs background contains every possible propagating direction, and because of stochastic fluctuation of the amplitudes of the HFRGWs, detection of the HFRGWs will be more difficult than that of the monochromatic plane GWs. In this case, can the HFRGWs be selected and measured by the EM detecting system? In particular, if both the HFRGWs have the same amplitude and frequency, but propagate along the exactly opposite directions (standing wave), will their effect be cancelled and nullified? We shall show that in our EM system the EM perturbation produced by the HFRGWs propagating along the positive and negative directions of the symmetrical axis (the $z$-axis) of the GB will be non-symmetric, and the physical effect generated by the HFRGWs propagating along other directions will be also quite different, even if they satisfy the resonant condition ($\omega_e = \omega_g$), and only the HFRGW component propagating along the positive direction of the symmetrical axis of the GB can generate an optimal resonant response. Thus our EM system would be very sensitive to the propagating directions as well as the frequencies of the HFRGWs

## 3. The coupling system of the fractal membranes and the Gaussian beam passing though a static magnetic field

Our EM system consists of the GB of a fundamental frequency mode [32] operating in the GHz band, a static magnetic field and the new-type fractal membranes [29-31]. It is well known [32] that the general form of the GB of a fundamental frequency mode is

$$\psi = \frac{\psi_0}{\sqrt{1+(z/f)^2}} \exp(-\frac{r^2}{W^2}) \exp\left\{i[(k_e z - \omega_e t) - \tan^{-1}\frac{z}{f} + \frac{k_e r^2}{2R} + \delta]\right\}, \qquad (4)$$

where $r^2 = x^2 + y^2$, $k_e = 2\pi/\lambda_e$, $f = \pi W_0^2/\lambda_e$, $W = W_0[1+(z/f)^2]^{1/2}$, $R = z + f^2/z$, $\psi_0$ is the amplitude of electric (or magnetic) field of the GB, $W_0$ is the minimum spot radius, $R$ is the curvature radius of the wave fronts of the GB at z, $\omega_e$ is the angular frequency, $\lambda_e$ is the EM wavelength, the z-axis is the symmetrical axis of the GB, and $\delta$ is an arbitrary phase factor. Different from Refs. [14, 15], here we choose the GB with the double transverse polarized electric modes (DTEM), and utilize the coupling effect between the fractal membrane in the GHz band and the GB passing through a static magnetic field (the fractal membranes can effectively reflect or transmit the photon flux in the GHz band [29-31]). In fact, the GBs with the DTEM exhibit more realizable modes, they have been extensively discussed and applied in the closed resonant cavities, open resonators and free space [32-35], including the standing-wave-type and traveling-wave-type GBs. Thus utilization of such GBs has more realistic significance. If the static magnetic field pointing along the y-axis is localized in the region $-l_1 \leq z \leq l_2$, setting $\tilde{E}_x^{(0)} = \psi = \psi_x$ and using divergenceless condition $\nabla \cdot \mathbf{E} = \frac{\partial \psi_x}{\partial x} + \frac{\partial \psi_y}{\partial y} = 0$ and $\tilde{\mathbf{B}}^{(0)} = -\frac{i}{\omega_e}\nabla \times \tilde{\mathbf{E}}^{(0)}$ (we use MKS units), then we have

$$\tilde{E}_x^{(0)} = \psi = \psi_x, \qquad \tilde{E}_y^{(0)} = \psi_y = -\int \frac{\partial \psi_x}{\partial x} dy = 2x(\frac{1}{W^2} - i\frac{k_e}{2R})\int \psi_x dy, \qquad \tilde{E}_z^{(0)} = 0,$$

$$\tilde{B}_x^{(0)} = \frac{i}{\omega_e}\frac{\partial \psi_y}{\partial z}, \qquad \tilde{B}_y^{(0)} = -\frac{i}{\omega_e}\frac{\partial \psi_x}{\partial z}, \qquad \tilde{B}_z^{(0)} = \frac{i}{\omega_e}(\frac{\partial \psi_x}{\partial y} - \frac{\partial \psi_y}{\partial x}), \tag{5}$$

and

$$\hat{B}^{(0)} = \begin{cases} \hat{B}_y^{(0)} & (-l_1 \leq z \leq l_2) \\ 0 & (z \leq -l_1 \text{ and } z > l_2) \end{cases}, \tag{6}$$

where the superscript 0 denotes the background EM fields, the notations ~ and ^ stand the time-dependent and static EM fields, respectively. For the high-frequency EM power flux (or in quantum language: photon flux), only non-vanishing average values of this with respect to time have an observable effect. From Eqs. (4) and (5), one finds

$$n_x^{(0)} = \frac{1}{\hbar\omega_e}\langle\frac{1}{\mu_0}(\tilde{E}_y^{(0)}\tilde{B}_z^{(0)})\rangle = \frac{1}{2\mu_0\hbar\omega_e}\text{Re}\left\{\psi_y^*[\frac{i}{\omega_e}(\frac{\partial \psi_x}{\partial y} - \frac{\partial \psi_y}{\partial x})]\right\} = f_x^{(0)}\exp(-\frac{2r^2}{W^2}), \tag{7}$$

$$n_y^{(0)} = -\frac{1}{\hbar\omega_e}\langle\frac{1}{\mu_0}(\tilde{E}_x^{(0)}\tilde{B}_z^{(0)})\rangle = \frac{1}{2\mu_0\hbar\omega_e}\text{Re}\left\{\psi_x^*[\frac{i}{\omega_e}(\frac{\partial \psi_y}{\partial x} - \frac{\partial \psi_x}{\partial y})]\right\} = f_y^{(0)}\exp(-\frac{2r^2}{W^2}), \tag{8}$$

$$n_z^{(0)} = -\frac{1}{\hbar\omega_e}\langle\frac{1}{\mu_0}(\tilde{E}_x^{(0)}\tilde{B}_y^{(0)}) - \frac{1}{\mu_0}(\tilde{E}_y^{(0)}\tilde{B}_x^{(0)})\rangle$$

$$= \frac{1}{2\mu_0\hbar\omega_e}\text{Re}\left\{\psi_x^*[\frac{i}{\omega_e}(\frac{\partial \psi_y}{\partial z})] + \psi_y^*[\frac{i}{\omega_e}(\frac{\partial \psi_x}{\partial z})]\right\} = f_z^{(0)}\exp(-\frac{2r^2}{W^2}), \tag{9}$$

where $\hbar\omega_e$ is the energy of single photon, $n_x^{(0)}$, $n_y^{(0)}$ and $n_z^{(0)}$ represent the average values of the background photon flux densities, in units of photons per second per square meter, propagating along the x-, y- and z-axes, respectively, the angular brackets denote the average over time, $f_x^{(0)}$, $f_y^{(0)}$ and $f_z^{(0)}$ are the functions of $\psi_0$, $W_0$, $\omega_e$, $r$ and $z$. Because of the non-vanishing $n_x^{(0)}$ and $n_y^{(0)}$, the GB will be asymptotically spread as $|z|$ increases (i.e., the irradiance surface of the GB spreads out in the + z and – z directions).

## 4. The EM resonant response to the HFRGWs

Using the electrodynamical equations in the curved spacetime

$$\frac{\partial}{\partial x^{\nu}}(\sqrt{-g}\, g^{\mu\alpha} g^{\nu\beta} F_{\alpha\beta}) = 0, \tag{10}$$

$$\nabla_{\alpha} F_{\mu\nu} + \nabla_{\nu} F_{\alpha\mu} + \nabla_{\mu} F_{\nu\alpha} = 0, \tag{11}$$

we can describe the EM perturbation produced by the HFRGWs in the EM system, where $F_{\mu\nu}$ is the EM field tensor, and $F_{\mu\nu} = F_{\mu\nu}^{(0)} + \tilde{F}_{\mu\nu}^{(1)}$, $F_{\mu\nu}^{(0)}$ and $\tilde{F}_{\mu\nu}^{(1)}$ represent the background and first-order perturbative EM fields respectively in the presence of the HFRGWs. Because of the weak field property of the HFRGWs, the perturbation methods will be valid. However, unlike plane monochromatic GWs, the amplitudes of the relic GW in Eq. (3) are not constant, in this case solving Eqs. (10) and (11) will often be difficult. In our case, fortunately, since $\nu_g = \omega_g / 2\pi = 10^{10}$ Hz and considering Eq. (3), the following equivalent relations would be valid provided $\omega_g \gg \dot{a}/a$, i.e.,

$$\frac{\partial}{\partial t} \to \mp i\omega_g, \quad \nabla \to i\mathbf{k}_g. \tag{12}$$

In this case the process of solving Eqs. (10) and (11) can be greatly simplified without excluding their essential physical features.

According to Eqs. (10), (11) and the equivalent relations, Eq. (12), the first-order perturbative EM fields generated by the direct interaction of the z-component of a certain "monochromatic wave," Eq. (3), with the static magnetic field $\hat{B}_y^{(0)}$ can be given by [14,16,18]

$$\tilde{E}_x^{(1)} = \frac{i}{2} A_{\oplus} \hat{B}_y^{(0)} k_g c(z+l_1)\exp[i(k_g z - \omega_g t)] + \frac{1}{4} A_{\oplus} \hat{B}_y^{(0)} c \exp[i(k_g z + \omega_g t)],$$

$$\tilde{B}_y^{(1)} = \frac{i}{2} A_{\oplus} \hat{B}_y^{(0)} k_g (z+l_1)\exp[i(k_g z - \omega_g t)] - \frac{1}{4} A_{\oplus} \hat{B}_y^{(0)} \exp[i(k_g z + \omega_g t)],$$

$$\tilde{E}_y^{(1)} = -\frac{1}{2} A_{\otimes} \hat{B}_y^{(0)} k_g c(z+l_1)\exp[i(k_g z - \omega_g t)] + \frac{i}{4} A_{\otimes} \hat{B}_y^{(0)} c \exp[i(k_g z + \omega_g t)],$$

$$\tilde{B}_x^{(1)} = \frac{1}{2} A_{\otimes} \hat{B}_y^{(0)} k_g (z+l_1)\exp[i(k_g z - \omega_g t)] + \frac{i}{4} A_{\otimes} \hat{B}_y^{(0)} \exp[i(k_g z + \omega_g t)], \tag{13}$$

where $A_\oplus, A_\otimes \approx A(k_g)/a$, $B(k_g)/a$ [see, Eq. (3)], $-l_1 \leq z \leq l_2$. Eq. (13) shows that the first-order perturbative EM fields have a space accumulation effect ($\propto z$) in the interacting region, this is because the GWs (gravitons) and EM waves (photons) have the same propagating velocity, so that the two waves can generate an optimum coherent effect in the propagating direction [16, 18]. From Eqs. (5), (6) and (13), the total EM field tensors in the presence of the HFRGW can be written as

$$F_{\mu\nu} = F_{\mu\nu}^{(0)} + \tilde{F}_{\mu\nu}^{(1)}$$

$$= \begin{pmatrix} 0 & \frac{1}{c}(\tilde{E}_x^{(0)} + \tilde{E}_x^{(1)}) & \frac{1}{c}(\tilde{E}_y^{(0)} + \tilde{E}_y^{(1)}) & 0 \\ -\frac{1}{c}(\tilde{E}_x^{(0)} + \tilde{E}_x^{(1)}) & 0 & -\tilde{B}_z^{(0)} & \hat{B}_y^{(0)} + \tilde{B}_y^{(0)} + \tilde{B}_y^{(1)} \\ -\frac{1}{c}(\tilde{E}_y^{(0)} + \tilde{E}_y^{(1)}) & \tilde{B}_z^{(0)} & 0 & -(\tilde{B}_x^{(0)} + \tilde{B}_x^{(1)}) \\ 0 & -(\hat{B}_y^{(0)} + \tilde{B}_y^{(0)} + \tilde{B}_y^{(1)}) & \tilde{B}_x^{(0)} + \tilde{B}_x^{(1)} & 0 \end{pmatrix}. \quad (14)$$

In our exemplar EM system we have chosen the GB power of $P = 10\,\text{W}$ and the static magnetic field of $\hat{B}_y^{(0)} = 3\,\text{T}$. In this case corresponding magnetic field amplitude of the GB is only $\tilde{B}^{(0)} \sim 10^{-5}\,\text{T}$, so the ratio of $\tilde{B}^{(0)}$ and the background static magnetic field $\hat{B}_y^{(0)}$ is roughly $\tilde{B}^{(0)}/\hat{B}_y^{(0)} \sim 10^{-5}$. In this case we have neglected the perturbation EM fields produced by the directed interaction of the HFRGW with the GB.

By using Eqs. (4)-(6), (13), (14) and the energy-momentum tensor $T_{\mu\nu}$ of the EM fields in the presence of the HFRGWs, we can calculate the first-order PPFs produced by the HFRGW. We shall focus our attention to the 01-component $\overset{(1)}{T}{}^{01}$ of the first-order perturbation, it expresses actually the x-component of the power flux density (Poynting vector) of the EM fields. Thus, the corresponding first-order PPF will be $c/\hbar\omega_e \overset{(1)}{T}{}^{01}$. In this case, the coherent synchro-resonance ($\omega_e = \omega_g$) between the perturbative EM fields, Eq. (13) and the GB, Eq. (4) and (5), can be expressed as the following first-order PPF density, i.e., x-component of the PPF is

$$n_x^{(1)} = \frac{c}{\hbar\omega_e}\langle T^{01(1)}\rangle_{\omega_e=\omega_g} = -\frac{c}{\mu_0\hbar\omega_e}\langle \tilde{F}_\alpha^{0(0)}\tilde{F}^{1\alpha(1)} + \tilde{F}_\alpha^{0(1)}\tilde{F}^{1\alpha(0)}\rangle_{\omega_e=\omega_g}$$

$$= \frac{1}{\hbar\omega_e}\langle \frac{1}{\mu_0}\tilde{E}_y^{(1)}\tilde{B}_z^{(0)}\rangle_{\omega_e=\omega_g} = \frac{1}{2\mu_0\hbar\omega_e}\text{Re}\left\{\tilde{E}_y^{(1)*}\left[\frac{i}{\omega_e}(\frac{\partial\psi_x}{\partial y} - \frac{\partial\psi_y}{\partial x})\right]\right\}_{\omega_e=\omega_g}$$

$$= -\frac{1}{\hbar\omega_e}\cdot\left\{\frac{A_\otimes \hat{B}_y^{(0)}\psi_0 k_g y(z+l_1)}{4\mu_0[1+(z/f)^2]^{1/2}(z+f^2/z)}\sin\left(\frac{k_g r^2}{2R} - \tan^{-1}\frac{z}{f}\right) + \frac{A_\otimes \hat{B}_y^{(0)}\psi_0 y(z+l_1)}{2\mu_0 W_0^2[1+(z/f)^2]^{3/2}}\right.$$

$$\left.\cos\left(\frac{k_g r^2}{2R} - \tan^{-1}\frac{z}{f}\right)\right\}\cdot\exp\left(-\frac{r^2}{W^2}\right) - \frac{1}{\hbar\omega_e}\left\{\left(1-\frac{4x^2}{W^2}\right)\frac{A_\otimes \hat{B}_y^{(0)}\psi_0 k_g(z+l_1)}{4\mu_0 R[1+(z/f)^2]^{1/2}}\cdot\left[F_1(y)\sin\left(\frac{k_g x^2}{2R} - \tan^{-1}\frac{z}{f}\right)\right.\right.$$

$$\left.+F_2(y)\cos\left(\frac{k_g x^2}{2R} - \tan^{-1}\frac{z}{f}\right)\right] + \left[\frac{2}{W^2} + \left(\frac{k_g^2}{R^2} - \frac{4}{W^4}\right)x^2\right]\cdot\frac{A_\otimes \hat{B}_y^{(0)}\psi_0(z+l_1)}{4\mu_0[1+(z/f)^2]^{1/2}}\left[F_1(y)\cos\left(\frac{k_g x^2}{2R} - \tan^{-1}\frac{z}{f}\right)\right.$$

$$\left.\left.-F_2(y)\sin\left(\frac{k_g x^2}{2R} - \tan^{-1}\frac{z}{f}\right)\right]\right\}\exp\left(-\frac{x^2}{W^2}\right), \tag{15}$$

where

$$F_1(y) = \int \exp(-\frac{y^2}{W^2})\cos(\frac{k_g y^2}{2R})dy,$$

$$F_2(y) = \int \exp(-\frac{y^2}{W^2})\sin(\frac{k_g y^2}{2R})dy, \tag{16}$$

are the quasi-probability integrals.

It is very interesting to compare $n_x^{(0)}$, Eq. (7), and $n_x^{(1)}$, Eq. (15). From Eqs. (5) and (7), we can see that $\tilde{E}_y^{(0)} = 0$ at the surface $x=0$, thus $n_x^{(0)}|_{x=0} = 0$; while numerical calculation shows that $n_x^{(1)}|_{x=0}$ has a non-vanishing observable value. This means that any photon measured by a detector (a tunable microwave receiver) from $n_x^{(1)}|_{x=0}$ will be a signal of the EM perturbation produced by the GW. Nevertheless, in the regions of $x \neq 0$, we have $n_x^{(0)} \neq 0$. At first sight $n_x^{(1)}$ will be swamped by the background $n_x^{(0)}$, so that $n_x^{(1)}$ has no observable effect in this regions. However, it will be shown that $n_x^{(1)}$ and $n_x^{(0)}$ propagate along the opposite directions in some local regions, and they have the *different rates of decay* [ $n_x^{(1)} \propto \exp(-\frac{r^2}{W^2})$ and $\exp(-\frac{x^2}{W^2})$, while $n_x^{(0)} \propto \exp(-\frac{2r^2}{W^2})$, see, Eqs. (7), (15) ]. Thus $n_x^{(1)}$ and $n_x^{(0)}$ can be separated by the special fractal membranes (see below), so that $n_x^{(1)}$ (signal), in principle, *would be observable*. The total PPF passing through a certain "typical receiving surface" $\Delta s$ at the *yz*-plane will be

$$N_x^{(1)} = \iint_{\Delta s} n_x^{(1)}|_{x=0}\, dydz. \tag{17}$$

Notice that $N_x^{(1)}$ is a unique non-vanishing photon flux passing through the surface i.e., number of photons per second (see, Fig. a). Equations (15) and (16) show that $n_x^{(1)}$ is an even function of the coordinates $x$, thus $n_x^{(1)}$ has the same propagating direction in the regions of $x > 0$ and $x < 0$; and at the same time, $n_x^{(1)}$ is an odd function of the coordinate y, so the propagating directions of $n_x^{(1)}$ are anti-symmetric to the regions of $y > 0$ and $y < 0$. (see, Fig. b, and such property ensured conservation of the total momentum of the PPF.) Considering the outgoing (and imploding, i.e., they go in both directions) property of $N_x^{(0)}$ in the region $z > 0$ (and $z < 0$) (this is a typical property of the GB [32]), it can be seen that $N_x^{(1)}$ and $N_x^{(0)}$ propagate along opposite directions in the regions of 1st ($x, y, z > 0$), 3rd ($x, y < 0, z > 0$), 6th ($x < 0, y > 0, z < 0$) and 8th ($x > 0$, $y, z < 0$) octants, while they have the same propagating directions in the regions of 2nd, 4th, 5th and 7th octants. (In Fig. b, we drew $N_x^{(1)}$ and $N_x^{(0)}$ in the 1st and 2nd octants.). For example, the +x directed PPS signal, $N_x^{(1)}$' (reflected by the fractal membranes in the y-z plane at the origin as in Fig. d) decays far less than the +x directed BPF $N_x^{(0)}$, so that the signal exceeds the BPF by the time they reach the detector located farther up on the +x axis. In our EM system example, all the parameters are chosen to exhibit values that can be realized in the proposed laboratory experiments:

(1) P=10W, the power of the GB. In this case, $\psi_0 \approx 1.26 \times 10^3 \text{ Vm}^{-1}$ for the GB of the spot radius $W_0 = 0.05 \text{ m}$.

(2) $\hat{B}_y^{(0)} = 3\,\text{T}$, the strength of the background static magnetic field.

(3) $0 \leq y \leq W_0$, $0 \leq z \leq 0.3\,\text{m}$, the integration region $\Delta s$ (the receiving surface of the PPF) in Eq. (17), i.e., $\Delta s \approx 10^{-2}\,\text{m}^2$.

(4) $l = l_2 + l_1 = 0.3\,\text{m}$, the interacting dimensions between the relic GW and the static magnetic field.

(5) $\nu_e = \nu_g = 10^{10}\,\text{Hz}$ and $A_\otimes = 2.00 \times 10^{-31}$ or $A_\otimes = 6.32 \times 10^{-31}$, they are typical orders of magnitude of the HFRGWs in the QIM [1-3] and in the PBBS [4-6], respectively. From Eqs. (15), (16) and (17), we obtain $N_x^{(1)} = 2.02 \times 10\,\text{s}^{-1}$ for the relic GW in the QIM and $N_x^{(1)} = 6.37 \times 10\,\text{s}^{-1}$ for the relic GW in the PBBS (see, Table 1). If the interacting dimension is increased to 6m, i.e., $l = l_2 + l_1 = 6\,\text{m}$, then the PPF (signal) produced by the HFRGWs in the QIM and in the PBBS may reach up to $4.04 \times 10^2\,\text{s}^{-1}$ and $1.27 \times 10^3\,\text{s}^{-1}$, respectively (see, table 1). The quantum picture of the above-mentioned process can be described as the resonant interaction of the photons with the gravitons in a background of virtual photons (the statistic magnetic field) as a catalyst [18,36], i.e., the inverse Gertsenshtein effect [13] involving elastic scattering of the gravitons to the photons in the background of virtual photons, which can greatly increase the interaction cross section between the photons and the gravitons. In other words the interaction may effectively change the physical behavior (e.g., propagating direction, distribution, polarization, and phase) of the partial photons in the local regions, and it does not require the resonant conversion of the gravitons to the photons, the latter corresponds to an extremely small conversion rate [37]. Consequently, even if the net increase of the photon number (the EM energy) of the entire EM system approaches zero, then one still might find the observable effect. In this case the requirements of relative parameters can be greatly relaxed, such properties may be very useful in order to display the very weak signal of the HFRGWs. In the case of astrophysical phenomenon, an analogous example is deflection of light (an EM wave beam) in a gravitational field, which causes the deflection of the propagating direction of the light ray, and although there is not any change of the photon number, but there is an observable effect. Of course in this process the interacting gravitational field is static (e.g., the gravitational field of the Sun). Thus

there is no the frequency resonant effect between the GWs and the EM waves and the space accumulation effect caused by the coherent interaction of the two kinds of waves in the propagating direction, but huge celestial gravitational fields compensated such shortcoming. In our system the change of the propagating directions of the partial photons in the local regions is caused by the GW, while the strong background static magnetic field as a catalyst and the resonant effect between the EM wave (the photon flux) and the GW (gravitons) compensate in part the weakness of the HFRGW. By the way, in our case even if the PPF signal equals 1000 s$^{-1}$ in a surface of 10$^{-2}$ m$^2$ area, there will still be a very small relative quantity of the BPF. For the GB with $\nu$=10$^{10}$ Hz, P=10W, $W_0$=5cm (the minimum spot radius of the GB), corresponding background photon flux is about ~10$^{23}$-10$^{24}$ s$^{-1}$ in the surface area, thus the share of the PPF in the BPF would be only ~10$^{-20}$-10$^{-21}$, and the PPF is not a net increase in photons, it is only the photons perturbed by the GW in the background photon flux (in our case the net increase of the photon number of the entire system approaches zero, but the PPF in some local regions would be non-vanishing). Moreover, because such PPF reflected by the special fractal membranes can be extracted from the BPF due to their very different decay rates, it would be observable and measurable.

## 5. The selective reflection of the perturbative photon flux.

It should be pointed out that because of the random property of the relic GWs, detection of the relic GWs will be more difficult than that of the continuous monochromatic plane GWs. However, we shall show that only the relic GW component propagating along the positive direction of the *z*-axis can generate optimal resonant response. It is true that for the relic GW components propagating along the *x-*, *y-* axes and negative direction of the *z*-axis, even if $\omega_g = \omega_e$, the PPFs produced by them will be much less than that generated by the relic GW component propagating along the positive direction of the *z*-axis; Thus the perturbations produced by the relic GW components propagating along the different directions cannot be counteracted for each other. In Fig. c we draw the symmetrical axis (the *z*-axis) of the Gaussian beam and the propagating directions $\mathbf{k}_g$ of the arbitrary component of the relic GWs.

In order to compare the PPFs generated by the different components of the HFRGW, we shall discuss the perturbations caused by the HFRGW's components propagating along some typical directions.

5-1. The PPFs generated by the HFRGW components propagating along different directions

(a). $\theta = 0$, i.e., the HFRGW component propagates along the positive direction of the z-axis. As is calculated the PPFs generated by the component can reach up to $4.04 \times 10^2 \, \text{s}^{-1}$ (the QIM) and $1.27 \times 10^3 \, \text{s}^{-1}$ (the PBBs) in a surface of $10^{-2} \, \text{m}^2$ area (see, table 1), respectively.

(b). $\theta = \pi$, i.e., the HFRGW component propagates along the negative direction of the z-axis. By using the similar means, one finds

$$n_x^{(1)} = -\frac{1}{\hbar \omega_e} \cdot \left\{ \frac{A_\otimes \hat{B}_y^{(0)} \psi_0 k_g y(l_2 - z)}{4\mu_0 [1+(z/f)^2]^{1/2}(z+f^2/z)} \sin\left(2k_g z + \frac{k_g r^2}{2R} - \tan^{-1}\frac{z}{f}\right) + \frac{A_\otimes \hat{B}_y^{(0)} \psi_0 y(l_2 - z)}{2\mu_0 W_0^2 [1+(z/f)^2]^{3/2}} \right.$$

$$\cos\left(2k_g z + \frac{k_g r^2}{2R} - \tan^{-1}\frac{z}{f}\right) \cdot \exp\left(-\frac{r^2}{W^2}\right) - \frac{1}{\hbar \omega_e} \left\{ \left(1 - \frac{4x^2}{W^2}\right) \frac{A_\otimes \hat{B}_y^{(0)} \psi_0 k_g (l_2 - z)}{4\mu_0 R [1+(z/f)^2]^{1/2}} \cdot \left[ F_1(y)\sin\left(2k_g z + \frac{k_g x^2}{2R} - \tan^{-1}\frac{z}{f}\right) \right.\right.$$

$$\left. + F_2(y)\cos\left(2k_g z + \frac{k_g x^2}{2R} - \tan^{-1}\frac{z}{f}\right)\right] + \left[\frac{2}{W^2} + \left(\frac{k_g^2}{R^2} - \frac{1}{W^4}\right)x^2\right] \cdot \frac{A_\otimes \hat{B}_y^{(0)} \psi_0 (l_2 - z)}{4\mu_0 [1+(z/f)^2]^{1/2}} \left[ F_1(y)\cos\left(2k_g z + \frac{k_g x^2}{2R} - \tan^{-1}\frac{z}{f}\right)\right.$$

$$\left.\left.- F_2(y)\sin\left(2k_g z + \frac{k_g x^2}{2R} - \tan^{-1}\frac{z}{f}\right)\right]\right\} \exp\left(-\frac{x^2}{W^2}\right) \quad (18)$$

Different from Eq. (15), each and all terms in Eq. (18) contain oscillating factor $2k_g z$. We emphasize that $2k_g z \approx 419z$ for the high-frequency relic GW of $\nu_g = 10^{10}$ Hz, namely, the factor $2k_g z$ will play major role in the region of the effective coherent resonance. In other words, the sign of $n_x^{(1)}$ will oscillate quickly and quasi-periodically change as the coordinate $z$ in the region increases. Thus the total effective PPF passing through a certain "typical receiving surface" will be much less than that generated by the relic GW component propagating along the positive direction of the z-axis, (see Eq. (15) and Table 2) (c). $\theta = \pi/2$, $\phi=0$, i.e., the propagating direction of the relic GW component is not only perpendicular to the symmetrical z-axis of the GB, but also perpendicular to the static magnetic field $\hat{B}_y^{(0)}$ directed along the y-axis. Here we assume that the dimension of the x-direction of $\hat{B}_y^{(0)}$ is localized in the region $-l_3 \leq x \leq l_4$. Utilizing the similar means the first-order perturbative EM fields generated by the direct interaction of the relic GW with the static magnetic field can be given by

$$\tilde{E}_y^{(1)} = \frac{i}{2} A_\oplus \hat{B}_y^{(0)} k_g \, c(x+l_3) \exp[i(k_g x - \omega_g t)] + \frac{1}{4} A_\oplus \hat{B}_y^{(0)} c \exp[i(k_g x + \omega_g t)],$$

$$\tilde{B}_z^{(1)} = \frac{i}{2} A_\oplus \hat{B}_y^{(0)} k_g \, (x+l_3) \exp[i(k_g x - \omega_g t)] - \frac{1}{4} A_\oplus \hat{B}_y^{(0)} \exp[i(k_g x + \omega_g t)],$$

$$\tilde{E}_z^{(1)} = -\frac{1}{2} A_\otimes \hat{B}_y^{(0)} \, k_g c(x+l_3) \exp[i(k_g x - \omega_g t)] + \frac{i}{4} A_\otimes \hat{B}_y^{(0)} c \exp[i(k_g x + \omega_g t)],$$

$$\tilde{B}_y^{(1)} = \frac{1}{2} A_\otimes \hat{B}_y^{(0)} k_g \, (x+l_3) \exp[i(k_g x - \omega_g t)] + \frac{i}{4} A_\otimes \hat{B}_y^{(0)} \exp[i(k_g x + \omega_g t)], \; (-l_3 \leq x \leq l_4). \quad (19)$$

In this case the coherent synchro-resonance ($\omega_e = \omega_g$) between the perturbative fields, Eq. (19), and the GB can be expressed as the following first-order PPF density, i.e.,

$$n_x^{(1)} = \frac{1}{\mu_0 \hbar \omega_e} \left[ \langle \tilde{E}_y^{(1)} \tilde{B}_z^{(0)} \rangle + \langle \tilde{E}_y^{(0)} \tilde{B}_z^{(1)} \rangle - \langle \tilde{E}_z^{(1)} \tilde{B}_y^{(0)} \rangle \right]_{\omega_e = \omega_g}, \quad (20)$$

where $\tilde{B}_y^{(0)}$ and $\tilde{B}_z^{(0)}$ are the y- and z- components of the magnetic filed of the GB, respectively, the angular brackets denote the average over time. Notice that we choose the GB of the transverse electric modes, so $\tilde{E}_z^{(0)} = 0$. By using the same method, we can calculate $n_x^{(1)}$, Eq. (20). For example, first term in Eq. (20) can be written as

$$\frac{1}{\mu_0 \hbar \omega_e} \langle \tilde{E}_y^{(1)} \tilde{B}_z^{(0)} \rangle_{\omega_e = \omega_g} = -\frac{1}{\hbar \omega_e} \left\{ \frac{A_\oplus \hat{B}_y^{(0)} \psi_0 k_g y(x+l_3)}{4\mu_0 \left[1+(z/f)^2\right]^{1/2} (z+f^2/z)} \sin\left( k_g (x-z) + \frac{k_g r^2}{2R} - \tan^{-1} \frac{z}{f} \right) \right.$$

$$\left. + \frac{A_\oplus \hat{B}_y^{(0)} \psi_0 y(x+l_3)}{2\mu_0 W_0^2 \left[1+(z/f)^2\right]^{3/2}} \cos\left( k_g (x-z) + \frac{k_g r^2}{2R} - \tan^{-1} \frac{z}{f} \right) \right\} \exp(-\frac{r^2}{W^2})$$

$$-\frac{1}{\hbar\omega_e}\left\{(1-\frac{4x^2}{W^2})\frac{A_\oplus \hat{B}_y^{(0)}\psi_0 k_g y(x+l_3)}{4\mu_0 R\left[1+(z/f)^2\right]^{1/2}}\left[F_1(y)\sin\left(k_g(x-z)+\frac{k_g x^2}{2R}-\tan^{-1}\frac{z}{f}\right)\right.\right.$$

$$\left.+F_2(y)\cos\left(k_g(x-z)+\frac{k_g x^2}{2R}-\tan^{-1}\frac{z}{f}\right)\right]+\left[\frac{2}{W^2}+(\frac{k_g^2}{R^2}-\frac{1}{W^4})x^2\right]\frac{A_\oplus \hat{B}_y^{(0)}\psi_0 y(x+l_3)}{4\mu_0\left[1+(z/f)^2\right]^{1/2}}$$

$$\left[F_1(y)\cos\left(k_g(x-z)+\frac{k_g x^2}{2R}-\tan^{-1}\frac{z}{f}\right)-F_2(y)\sin\left(k_g(x-z)+\frac{k_g x^2}{2R}-\tan^{-1}\frac{z}{f}\right)\right]\bigg\}\exp(-\frac{x^2}{W^2}).$$

$$(-l_3 \leq x \leq l_4) \tag{21}$$

It can be shown that calculation for the 2nd and 3rd terms in Eq. (20) is quite similar to first term, and they have the same orders of magnitude, we shall not repeat it here. Notice that unlike $n_x^{(1)}$ produced by the relic GW component propagating along the positive direction of the $z$-axis [see, Eq. (15)], the phase functions in Eq. (21) contain oscillating factor $k_g(x-z)$, and because it is always possible to choose $l_2 + l_1 \gg l_4 + l_3$, i.e., the dimension of the $z$-direction of $\hat{B}_y^{(0)}$ is much larger than its $x$-direction dimension. Thus, the PPF expressed by Eq. (15) will be much larger than that repressed by Eq. (21) (see, Table 2).

(d) $\theta = \pi/2$, $\phi = \pi/2$, i.e., the relic GW component propagates along the y-axis, which is parallel with the static magnetic field $\hat{B}_y^{(0)}$.

According to the Einstein-Maxwell equations of the weak field, then the perturbation of the GW to the static magnetic field vanishes [16, 18], i.e.,

$$n_x^{(1)} = 0. \tag{22}$$

It is very interesting to compare $n_x^{(1)}$ in Eqs. (15), (18), (21) and (22), as is shown that although they all represent the PPFs propagating along the $x$-axis, their physical behaviors are quite different. In the case of $\theta = \phi = \pi/2$, $n_x^{(1)} = 0$, Eq. (22); when $\theta = \pi$ and $\theta = \pi/2$, $\phi = 0$, the PPFs contain the oscillating factors $2k_g z$ and $k_g(x-z)$, respectively [see Eqs. (18) and (21)]. Only under the condition $\theta = 0$, the PPF, Eq. (15), does not contain any oscillating factor, but only slow enough variation function in the z direction. This means that $n_x^{(1)}$ produced by the relic GW component propagating along the positive direction of the *z*-axis, has the best space accumulation effect (see Table 2). Thus our EM system would be very sensitive to the propagating directions of the relic GWs. In other words the EM system has a strong selection capability to the resonant components from the stochastic relic GW background.

5-2. The separation of the PPFs (signal) from the BPFs

In recent years new types of fractal membranes have been successfully developed [29-31]. Firstly, these fractal membranes can provide nearly total reflection for the EM waves (photon flux) with certain frequencies in the GHz band; at the same time, they can provide nearly total transmission for the photon fluxes with other frequencies in the GHz band. Secondly, the photon fluxes reflected and transmitted by the fractal membranes can keep their strength *invariant* within the distance of one meter from the fractal-membrane's surface. Thirdly, such frequencies can be regulated in the GHz band. Since $N_x^{(1)}$ (signal) and $N_x^{(0)}$ (background) propagate along the negative and positive directions of the x-axis in the first octant (the region of $x, y, z > 0$), respectively, i.e., $N_x^{(1)}$ propagates along the direction toward the fractal membrane, while $N_x^{(0)}$ propagates along the direction away from the fractal membrane (see Fig. b). Using the reflecting fractal membrane with its plane normal to the x-axis, it will reflect *only* $N_x^{(1)}$ and *not* $N_x^{(0)}$. Once $N_x^{(1)}$ is reflected (defined as $N_x^{(1)'}$) it will have the same propagating direction as $N_x^{(0)}$. However, after $N_x^{(1)}$ is reflected, it can keep its strength *invariant* within one meter distance from the fractal membrane [29, 30], while $N_x^{(0)}$ decays as the typical Gaussian decay rate $\exp(-\frac{2r^2}{W^2})$ [see, Eq. (7)], then the ratio $N_x^{(1)'}/N_x^{(0)}$ (the signal-to-background noise ratio) would be larger than one in the whole region of $0.42\text{m} \leq x \leq 1\text{m}$ (see, Table 3, *x* is the distance from the detector to the fractal membrane). Table 3 shows that the BPF $N_x^{(0)}$ is much larger than the PPF $N_x^{(1)'}$ in the region $0 < x < 42\,\text{cm}$, while the $N_x^{(0)}$ and $N_x^{(1)'}$ have the same order of magnitude at $x = 42\,\text{cm}$, and $N_x^{(1)'}$ will be larger than $N_x^{(0)}$ in the region of $42\,\text{cm} < x < 100\,\text{cm}$ (i.e., where the signal-to-background noise ratio $N_x^{(1)'}/N_x^{(0)}$ would be larger than one). In particular, the BPF $N_x^{(0)}$ will be reduced to $\sim 10^{-6}\,\text{s}^{-1}$ at $x = 70\,\text{cm}$ while $N_x^{(1)}$' can keep the strength of $\sim 10^3\,\text{s}^{-1}$ at the same position [29, 30]. Thus it is possible to obtain an almost pure PPF (signal) at this position.

5-3. The thermal noise and the EM noise.

Besides the background noise issue just mentioned, the thermal noise and the possible external EM noise issue should be studied. Because the frequency of the PPF (signal) is roughly $10^{10}\,\text{Hz}$, if the system is cooled down to $kT < \hbar\omega_e$ ( $k$ is Boltzmann's constant, $\omega_e = 2\pi\nu_e$, $\nu_e = 10^{10}\,\text{Hz}$ ), i.e., $T < \hbar\omega_e / k \sim 0.48\,\text{K}$, then the frequency $\nu_m$ of the thermal photons will be less than the $\nu_e$ of the PPF. If the apparatus is kept to a lower temperature, e.g., $T < 0.048\,\text{K}$ (this is well within the current technology), then we have $\nu_m \approx 10^{-2}\nu_e$. Thus the difference in the frequency band for such two kinds of photons would be very great, i.e., the signal photon flux and the thermal photons can be easily distinguished. In other words, practically speaking there are *no thermal photons* at $10\,\text{GHz}$, and in this way the thermal noise can be suppressed as long as the EM detector can select the correct frequency.

For the possible external EM noise sources, using a Faraday cage or shielding covers made from such fractal membranes [29-31] would be very effective. Once the EM system is isolated from the outside world by the Faraday cage, and such background noise, thermal noise and possible dielectric dissipation (using a vacuum operation) can be effectively suppressed, one would obtain good environment of the signal-to-noise ratio.

Also, it should be pointed out that superposition of the relic GWs stochastic components will cause the fluctuation of the PPFs, even if such "monochromatic components" all satisfy the frequency resonant condition ( $\omega_e = \omega_g$ ). However, Eqs. (15), (18), (21) and (22) show that the metric perturbation only influences the strength fluctuation of the PPFs and does not influence the "direction resonance". That is, it does not influence the selection capability of the EM system to the propagating directions of the relic GWs, and it does not influence average effect over time of the

PPFs.

If the PPFs mentioned earlier cannot be effectively displayed and detected, then one could utilize a static magnetic field of 3T, having a cross-section of $3\times10^{-2}\,\text{m}^2$ and length of 100m. In this case the terminal microwave receiver's sensitivity would only need to detect the PPFs of $N_x^{(1)} \sim 6.73\times10^3\,\text{s}^{-1}$ for the HFRGWs in the QIM and $N_x^{(1)} \sim 2.12\times10^4\,\text{s}^{-1}$ for the HFRGWs in the PBBs, respectively.

Finally, it should be pointed out that the values of the PPFs discussed in the present paper depends on the strength of the HFRGWs in the GHz band expected by the QIM and other relevant string cosmology scenarios (e.g., see Refs, [1-6]). Because such models and scenarios are somewhat controversial, we cannot know in advance of our experiment whether or not these models and scenarios might have fatal flaws. If the real HFRGWs in the GHz band are much less than the magnitude expected by such models and scenarios, even if the required conditions can be satisfied and one still cannot to detect and measure such HFRGWs, then the HFRGW models will be suspect. Thus, this scheme might provide an indirect way to test such models and scenarios, that is, as suggested by Brustein et al.[38], a null experiment would be valuable. In any event, the HFGW generator and detector experiment described in Ref. [39] will prove the concept of the present detector. More detailed research into this subject remains to be done.

## Acknowledgments

This work is supported by the National Basic Research Program of China under Grant No.2003 CB 716300, the National Natural Science Foundation of China under Grant 10575140, the Nature Science Foundation of Chongqing under Grant 8562, GRAVWAVE ® LLC and Transportation Science Corporation of USA.

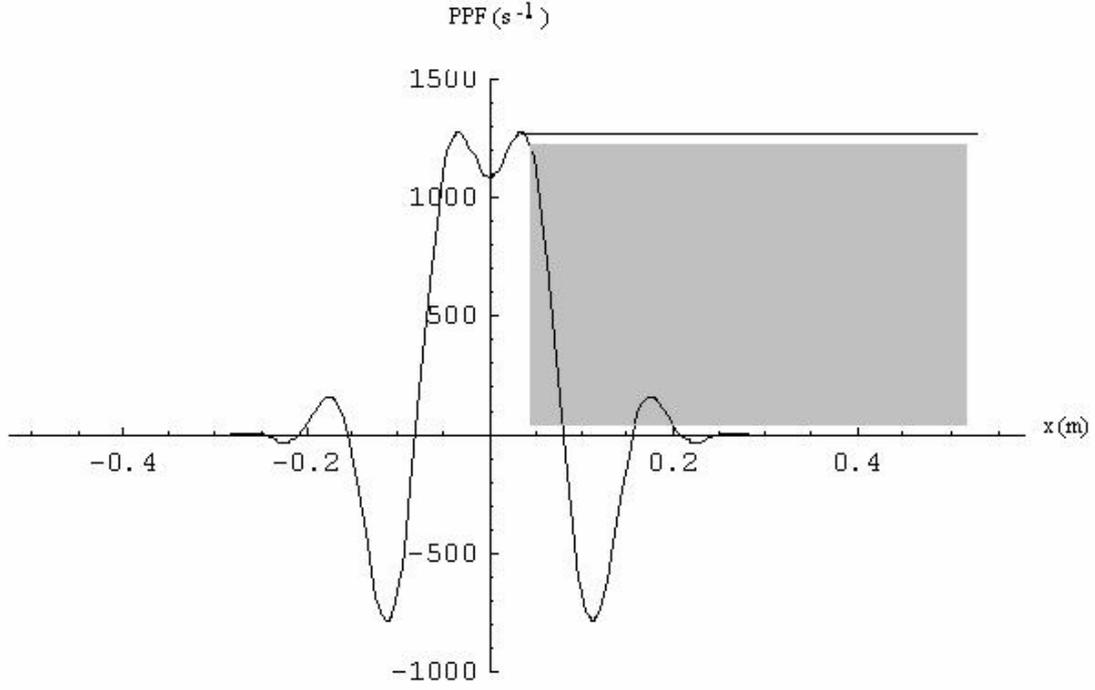

**Figure a.** The perturbative photon flux $N_x^{(1)}$ generated by the HFRGW in the PBBS. $\left|N_x^{(1)}\right| = \left|N_x^{(1)}\right|_{max} = 1.27 \times 10^3$ s$^{-1}$ at $x = \pm 3.16$cm, we take note of that the background photon flux $N_x^{(0)}\big|_{x=0} = 0$ [see, Eqs.(5) and (7)] while $N_x^{(1)}\big|_{x=0} = 1.15 \times 10^3$ s$^{-1}$, which would be an observable value, and $N_x^{(1)}$ and $N_x^{(0)}$ propagate along opposite directions in the first octant.

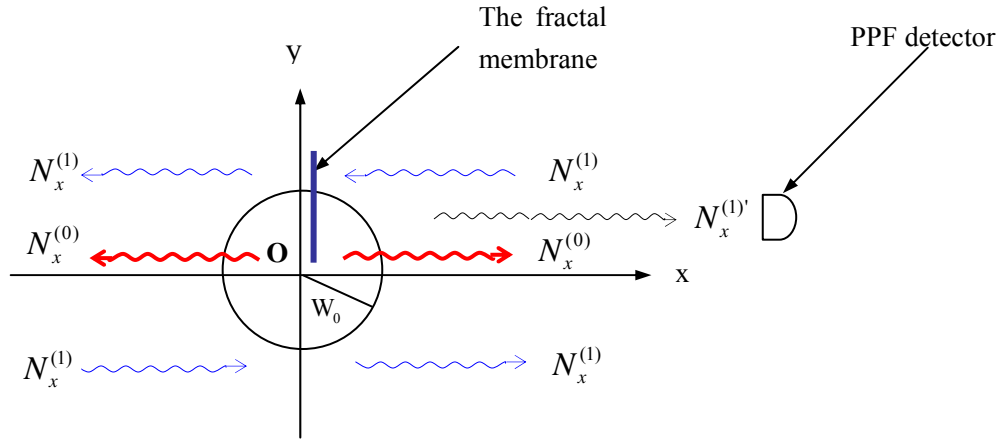

**Figure b.** $N_x^{(0)}$, $N_x^{(1)}$ and $N_x^{(1)'}$ in the 1st and 2nd octants. After $N_x^{(1)}$ is reflected by the fractal membrane, (i.e., $N_x^{(1)'}$ in the 1st octant), $N_x^{(1)'}$ and $N_x^{(0)}$ will have the same propagating direction. However, $N_x^{(1)'}$ can keep its strength invariant within one meter to the membrane (see, e.g., Refs. [29,30]), while $N_x^{(0)}$ decays as the typical Gaussian decay rate $\exp(-2r^2/W^2)$ [see, Eq. (7)], then the ratio $N_x^{(1)'}/N_x^{(0)}$ would be larger than one in the whole region of $0.42\,\mathrm{m} < x < 1\,\mathrm{m}$, although $N_x^{(0)} \gg N_x^{(1)'}$ in the region of $0 < x < 0.42\,\mathrm{m}$, here the resonant component of the GW propagates along the *z*-axis.

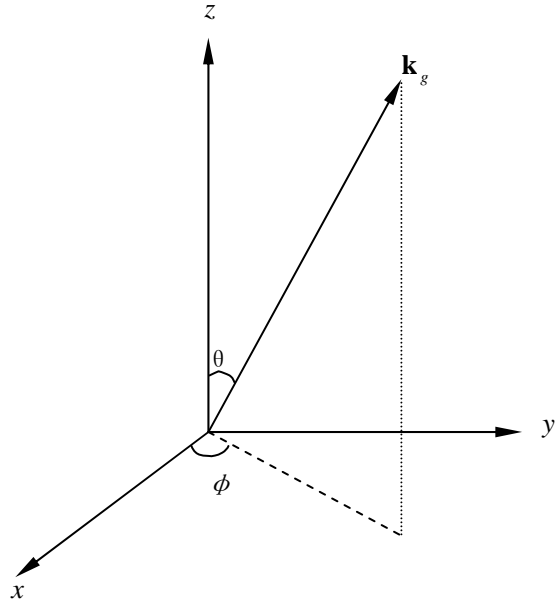

**Figure c**. The *z*-axis is the symmetrical axis of the Gaussian beam, $\mathbf{k}_g$ represents the instantaneous propagating direction of the arbitrary component of the relic GW

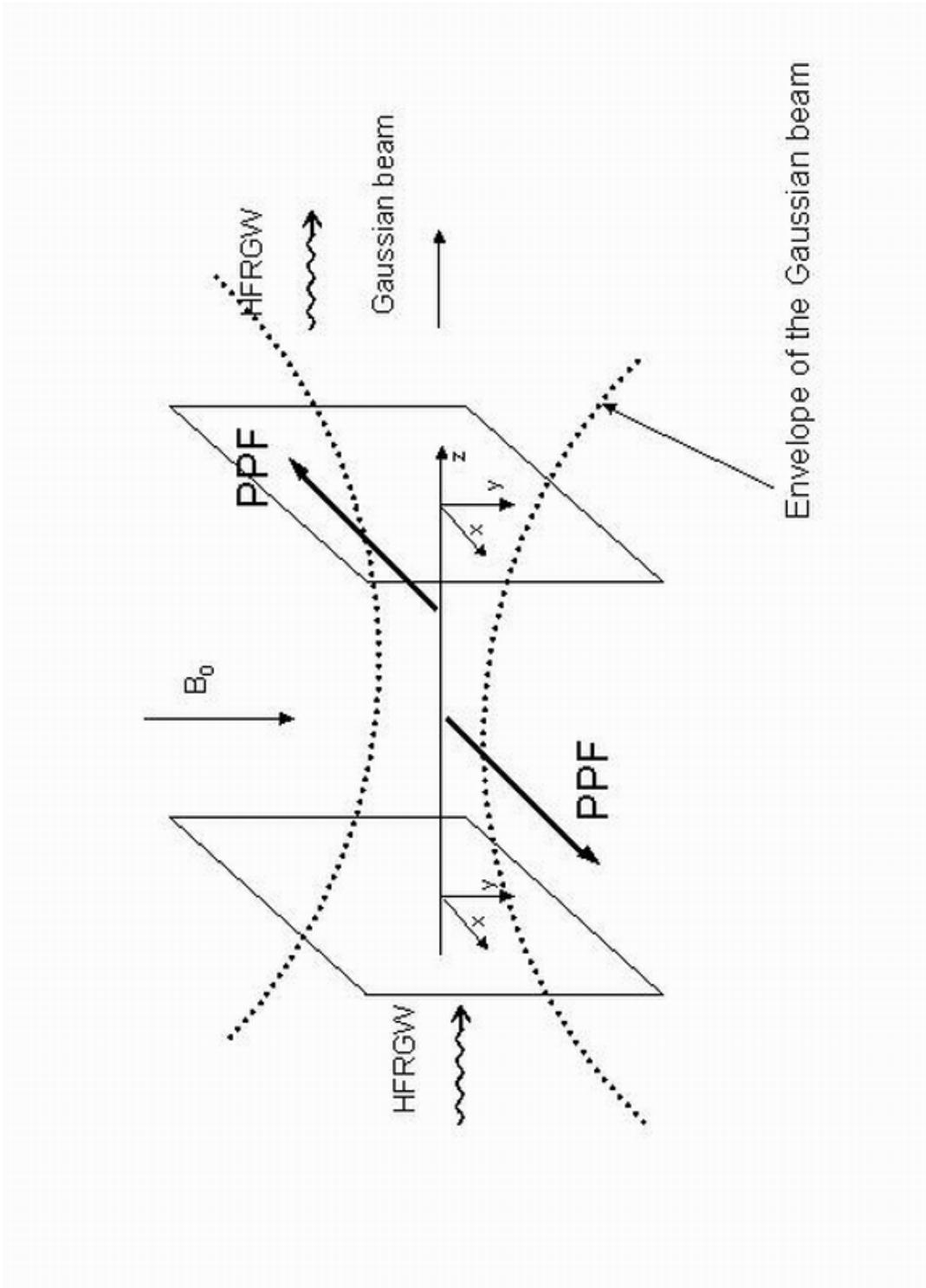

**Figure d**  This schema is lateral view of Figure b, propagating directions of x-components $N_x^{(1)}$ of the PPF are opposite to each other in the regions of y>0 and y<0. After $N_x^{(1)}$ is reflected by the fractal membrane, it will has much less decay rate than the BPF $N_x^{(0)}$, which has a typical Gaussian decay rate.

**Table 1.** The PPFs and relevant parameters. Here $\Omega_{gw}$ is the normalized energy density of the relic GWs, $\Omega_{gw} \sim h^2(\frac{\nu}{\nu_H})^2$ (see, e.g., Ref. [9]), so $h \sim \frac{\nu_H}{\nu}(\Omega_{gw})^{\frac{1}{2}}$ is the metric perturbation, $\nu_H \approx 2.00 \times 10^{-18}$ Hz is the Hubble frequency. The peak values of $\Omega_{gw}$ in the QIM [1-3] and in the PBBS [4-6] would be $\sim 10^{-6} - 10^{-5}$, respectively. The resolution of tiny difference between the PPFs generated by the relic GWs in the QIM and in the PBBS would be a significant subject for cosmological research.

| | $\nu$ (Hz) | $\Omega_{gw}$ | $h$ | $l$ (m), the interacting dimensions | The PPFs (s$^{-1}$) |
|---|---|---|---|---|---|
| The relic GW in the QIM | $10^{10}$ | $10^{-6}$ | $2.00 \times 10^{-31}$ | 0.30 | $2.02 \times 10$ |
| | | | | 6.00 | $4.04 \times 10^2$ |
| The relic GW in the PBBS | $10^{10}$ | $10^{-5}$ | $6.32 \times 10^{-31}$ | 0.30 | $6.37 \times 10$ |
| | | | | 6.00 | $1.27 \times 10^3$ |

**Table 2.** The PPFs generated by the resonant relic GW components propagating along the different directions, here $\hat{B}^{(0)} = 3\,\text{T}$, $A_\otimes$, $A_\oplus \sim 6.63 \times 10^{-31}$, $\nu_g = 10^{10}$ Hz, $l_2 + l_1 = 6\,\text{m}$

| Propagating directions of the resonant components of the relic GW | PPFs (s$^{-1}$) |
|---|---|
| z | $1.27 \times 10^3$ |
| -z | $3.16 \times 10$ |
| x | $6.30 \times 10^{-1}$ |
| y | 0 |

**Table 3.** Comparison of the PPF reflected by the fractal membrane and the BPF in the $x$-direction, here $\hat{B}^{(0)} = 3\,\text{T}$, $A_\otimes, A_\oplus \sim 6.63 \times 10^{-31}$, $\nu_g = 10^{10}\,\text{Hz}$, $l_2 + l_1 = 6\,\text{m}$. The PPF $N_x^{(1)}$ reflected (defined as $N_x^{(1)}$') by the fractal membrane can keep its strength invariant within one meter distance from the membrane[29-31], while the BPF $N_x^{(0)}$ decays as $\exp(-2r^2/W^2)$ [see Eq.(7)]. Thus the $N_x^{(1)}$' and $N_x^{(0)}$ would have the same order of magnitude at $x = 42\,cm$, and the terminal microwave receiver is possible to obtain an almost pure PPF(signal) at $x = 70\,cm$.

| The distance to the fractal membrane(cm) | 3.16 | 12 | 42 | 70 |
|---|---|---|---|---|
| $N_x^{(0)}$ (s$^{-1}$), the BPF | $2.44 \times 10^{24}$ | $2.16 \times 10^{22}$ | $1.27 \times 10^3$ | $1.97 \times 10^{-6}$ |
| $N_x^{(1)}$'(s$^{-1}$), the PPF reflected by the fractal membrane (3.16cm<$x$<100cm) | | | $1.27 \times 10^3$ | |